\documentstyle[epsfig,epsf,rotate,a4,12pt]{article}
\epsfverbosetrue

\def\nuebar{\bar{\nu_e}}
\def\nue{\nu_e}
\def\munu{\mu_{\nu}}
\def\gammanu{\Gamma_{\nu}}
\def\dm2{\rm{\Delta m^2}}

\def\s2tw{\rm{ sin ^2 \theta _W }}
\def\am241{\rm{ ^{241} Am }}
\def\u238{\rm{ ^{238} U }}
\def\th232{\rm{ ^{232} Th }}
\def\k40{\rm{ ^{40} K }}

\begin{document}

\begin{flushright}
   {\bf AS-TEXONO/02-01}\\
        \today 
\end{flushright}

\begin{center}
\Large
\bf{
Research Program of the TEXONO Collaboration:\\
Status and Highlights\\
}
\end{center}

\begin{center}
\large
Henry Tsz-King Wong~$^{\alpha ,}$\footnote{Contact Person:
htwong@phys.sinica.edu.tw}
and 
Jin Li~$^{\beta}$
\end{center}
\begin{flushleft}
\large
$^{\alpha}$ Institute of Physics, Academia Sinica, Taipei 11529, Taiwan\\
$^{\beta}$ Institute of High Energy Physics,
Chinese Academy of Science, \\
\hspace*{2cm} Beijing 100039, China\\
\end{flushleft}


\vspace*{0.5cm}

\begin{center}
\large
{\bf Abstract}
\normalsize
\end{center}

This article reviews
the research program and efforts for the
TEXONO Collaboration among scientists
from Taiwan and China.
These include reactor-based neutrino physics
at the Kuo-Sheng Power Plant in
Taiwan as well as various R\&D efforts 
related to the various experimental techniques
in neutrino and astro-particle physics. 

\vfill

\begin{center}
\it{Invited Talk at the 
1st NCTS Workshop on Astro-Particle Physics,\\
Kenting, Taiwan, Dec 2001.}
\end{center}

\clearpage

\section{Introduction and History}

The 
TEXONO\footnote{{\bf T}aiwan {\bf EX}periment {\bf O}n {\bf N}eutrin{\bf O}}
Collaboration\cite{texono} has been built up since 1997 to
initiate and pursue an experimental
program in Neutrino and Astroparticle Physics\cite{start}.
By the end of  2001, the Collaboration comprises
more than 40 research scientists from
major institutes/universities
in Taiwan (Academia Sinica$^{\dagger}$, Chung-Kuo
Institute of Technology, Institute of Nuclear
Energy Research, National Taiwan University, National Tsing Hua
University, and Kuo-Sheng Nuclear Power Station),
China (Institute of High Energy Physics$^{\dagger}$,
Institute of Atomic Energy$^{\dagger}$,
Institute of Radiation Protection,
Nanjing University, Tsing Hua University)
and the United States (University of Maryland),
with AS, IHEP and IAE (with $^{\dagger}$)
being the leading groups.
It is the first research collaboration 
of this size and magnitude.
among Taiwanese and Chinese scientists 
from major research institutes.

The research program\cite{program} is based on the
the unexplored and unexploited theme of adopting
detectors with
high-Z nuclei, such as solid state device and
scintillating crystals,
for low-energy low-background experiments
in Neutrino and Astroparticle Physics\cite{prospects}.
The ``Flagship'' program\cite{expt} is 
a reactor neutrino experiment
to study low energy neutrino
properties and interactions.
It is the first particle physics experiment
performed in Taiwan where local scientists
are taking up  major roles and responsibilities
in all aspects of its operation:
conception, formulation, design, prototype
studies, construction, commissioning,
as well as data taking and analysis.

In parallel to the reactor experiment,
various R\&D efforts coherent with the
theme are initiated and pursued. Subsequent
sections give the details and status of the program.

\section{Kuo-Sheng Neutrino Laboratory}

\begin{figure}
\center
\epsfig{figure=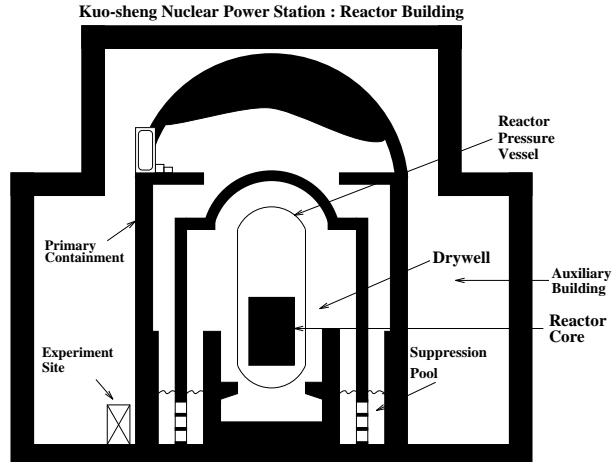,height=8cm,angle=270}
\caption{
Schematic side view, not drawn to scale,
of the Kuo-sheng Nuclear Power Station
Reactor Building,
indicating the experimental site.
The reactor core-detector distance is about
28~m.
}
\label{ksnpssite}
\end{figure}

The ``Kuo-Sheng Neutrino Laboratory''
is located at a distance of 28~m from the core \#1
of the Kuo-Sheng Nuclear Power Station (KSNPS)
at the northern shore of Taiwan\cite{expt}.
A schematic view is depicted in Figure~\ref{ksnpssite}.

\begin{figure}
\center
\epsfig{figure=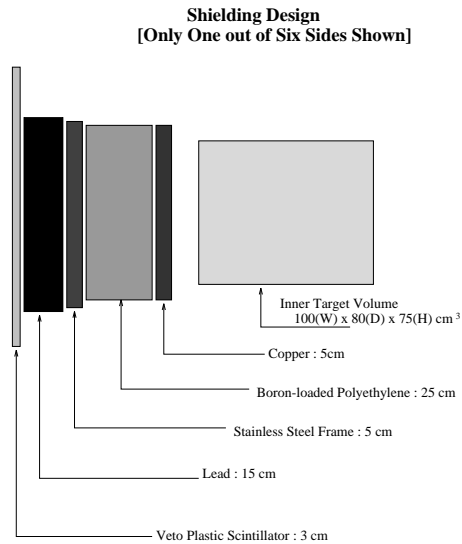,height=2.7in,angle=270}
\caption{
Schematic layout of the inner target space,
passive shieldings and cosmic-ray veto panels.
The coverage is 4$\pi$ but only one face
is shown.
}
\label{shield}
\end{figure}

A multi-purpose ``inner target'' detector space of
100~cm$\times$80~cm$\times$75~cm is
enclosed by 4$\pi$ passive shielding materials
cosmic-ray veto scintillator panels,
the schematic layout of  which is shown
in Figure~\ref{shield}
The shieldings provide attenuation
to the ambient neutron and gamma background,
and are made up of, inside out,
5~cm of OFHC copper, 25~cm of boron-loaded
polyethylene, 5~cm of steel and 15~cm of lead.

Different detectors can be placed in the
inner space for the different scientific goals.
The detectors will be read out by a versatile
electronics and data acquisition systems\cite{eledaq}
based on a 16-channel, 20~MHz, 8-bit 
Flash Analog-to-Digital-Convertor~(FADC)  module.
The readout allows full recording of all the relevant pulse
shape and timing information for as long as several ms
after the initial trigger.
The reactor laboratory is connected via telephone line to
the home-base laboratory at AS, where remote access 
and monitoring are performed regularly. Data are stored
and accessed in a multi-disks array with a total
of 600~Gbyte memory via IDE-bus in PCs.

It is recognized recently\cite{sensit} that
due to the uncertainties in the modeling of the
low energy part of the reactor neutrino spectra,
experiments to measure 
Standard Model neutrino-electron
cross sections with reactor neutrinos
should focus on higher electron recoil
energies (T$>$1.5~MeV), as with (b), while
neutrino magnetic moment searches should base
on measurements  with T$<$100~keV.

Accordingly, data taking for Period I Reactor ON/OFF
has started in July 2001 and will continue till
March 2002.
Two detector systems are running in parallel 
using the same data acquisition system 
but independent triggers:
(a) an Ultra Low Background High Purity Germanium (ULB-HPGe),
with a fiducial mass of 1.06 kg,
and (b) 46~kg of CsI(Tl) crystal scintillators.
The target detectors are housed in a nitrogen
environment to prevent background events due to the
diffusion of the radioactive radon gas.

\subsection{Germanium Detector}

\begin{figure}
\center
\epsfig{figure=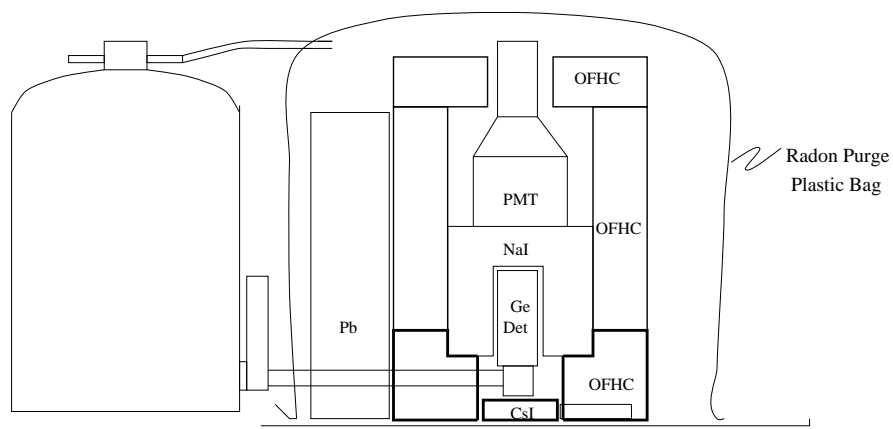,height=12cm,angle=270}
\caption{
Schematic drawings of the ULB-HPGe
detector with its anti-Compton scintillators
and passive shieldings.
}
\label{inhpge}
\end{figure}

As depicted in Figure~\ref{inhpge},
the ULB-HPGe is surrounded by NaI(Tl) and CsI(Tl) crystal
scintillators as anti-Compton detectors, and the whole set-up
is further enclosed by another 3.5~cm of OFHC copper and
lead blocks.

\begin{figure}
\center
\epsfig{figure=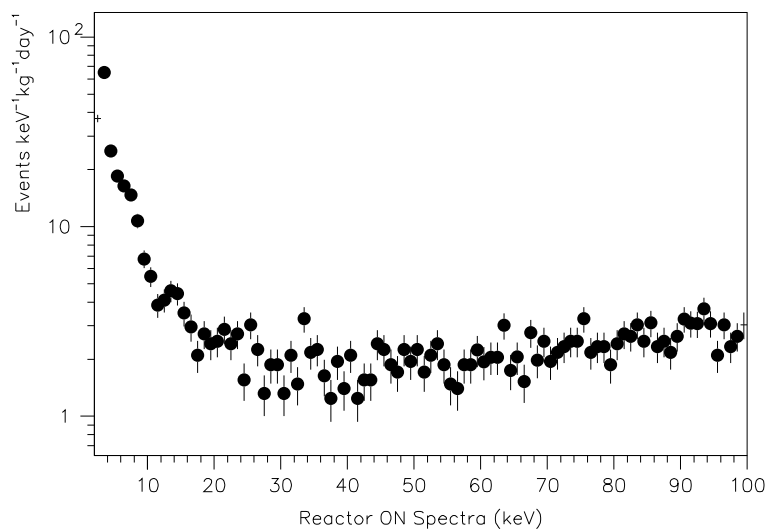,height=7cm}
\caption{
The measured spectrum from the ULB-HPGe, 
after cuts of cosmic and anti-Compton vetos, 
during 12.2~days of reactor ON data taking.
}
\label{gespec}
\end{figure}

The measured spectrum, after cuts of cosmic and
anti-Compton vetos, 
during 12.2~days of reactor ON data taking is displayed 
in Figure~\ref{gespec}.
Background (order of 1~keV$^{-1}$kg$^{-1}$day$^{-1}$)
and threshold (5~keV) levels comparable to underground Dark Matter
experiment has been achieved on site. 
Additional cuts based on pulse shape 
and timing information
are expected to further reduce the background level
at low energy.
It is expected the
data taken in Period I would allow us to achieve world
level sensitivities in 
$\nuebar$ magnetic moments ($\munu$)\cite{vogelengel}, 
and therefore indirectly, 
radiative lifetimes ($\gammanu$)\cite{raffelt}. 
These are the lowest threshold data so far for
reactor neutrino experiments, and therefore allow the studies
of more speculative topics, like $\munu$ and $\gammanu$
for $\nue$ from reactors, possible nuclear cross-sections, as
well as anomalous neutrino interactions.

\subsection{Scintillating CsI(Tl) Crystals}

The potential merits of crystal scintillators
for low-background low-energy experiments were
recently discussed\cite{prospects}.

\begin{figure}
\center
\epsfig{figure=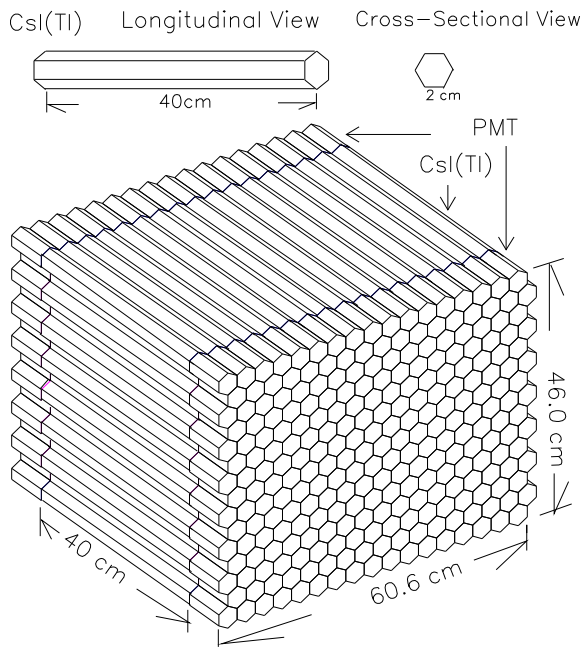,height=8cm}
\caption{
Schematic drawings of the CsI(Tl)
target configuration. A total of 23 modules (46~kg)
is installed for Period I.
}
\label{incsi}
\end{figure}

The CsI(Tl) detector
system is displayed in Figure~\ref{incsi}.
Each crystal module
is 2~kg in mass and 
consists of a hexagonal-shaped cross-section with 2~cm
side and a length 40~cm. 
The first sample is with
two 20~cm crystals glued optically
at one end to form a module (L20+20).
Techniques to grow CsI(Tl) mono-crystal
of length 40~cm (L40), the longest in the world for
commercial production,
have been developed and are deployed in
the production for subsequent batches.
The light output are read out
at both ends  by custom-designed 29~mm diameter
photo-multipliers (PMTs) with low-activity glass. 
The sum and difference of the PMT signals gives information
on the energy and the longitudinal position of
the events, respectively.

\begin{figure}
\centerline{
\epsfig{file=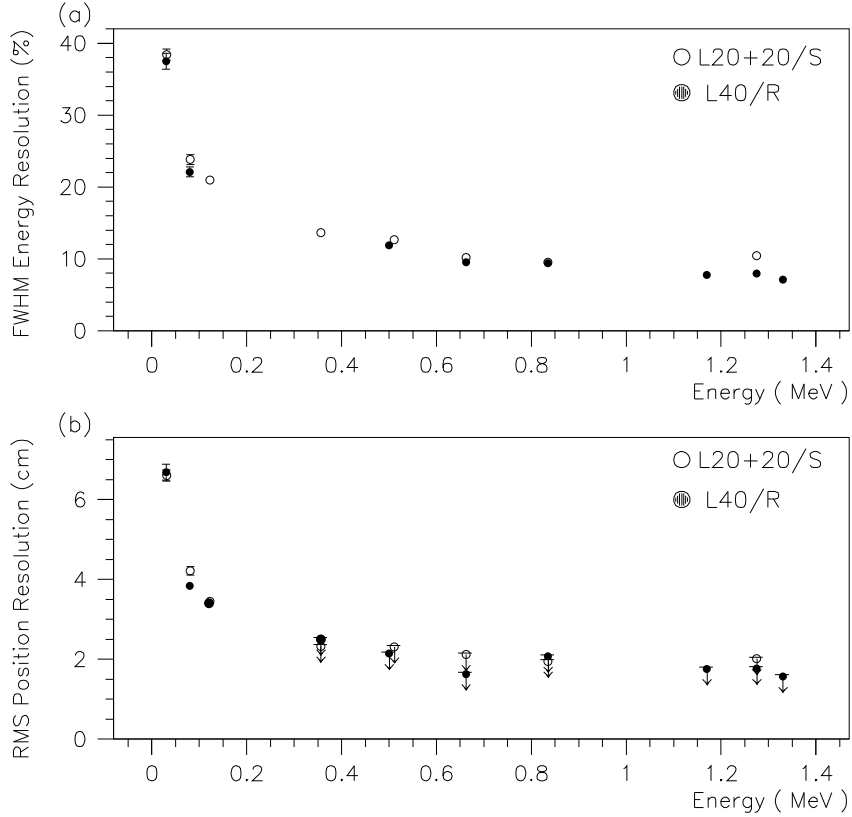,width=12cm}
}
\caption{
The variation of
(a) FWHM energy resolution
and (b) RMS position resolution
with energy
for the CsI(Tl) crystal modules.
Only upper limits are shown
for the higher energy points in (b)
since the events are not localized.
}
\label{csireso}
\end{figure}

Extensive measurements on the crystal prototype
modules have been performed\cite{proto}.
The energy and spatial resolutions as
functions of energy are  depicted
in Figure~\ref{csireso}. 
The energy is defined by the total
light collection  $\rm{Q_1 + Q_2}$.
It can be seen that
a $\sim$10\% FWHM energy 
resolution is achieved at 660~keV.
The detection threshold (where signals
are measured at both PMTs) is $<$20~keV.
The longitudinal position can be obtained
by considering the variation of the ratio
$\rm { R = (  Q_1 - Q_2 ) / (  Q_1 + Q_2 ) }$
along the crystal.
Resolutions of $\sim$2~cm and $\sim$3.5~cm at
660~keV and 200~keV, respectively, have been demonstrated.

In addition, CsI(Tl) provides powerful
pulse shape discrimination capabilities
to differentiate $\gamma$/e from $\alpha$ events,
with an excellent separation of $>$99\% above 500~keV.
The light output for $\alpha$'s in CsI(Tl) is quenched
less than that in liquid scintillators.
The absence of multiple $\alpha$-peaks 
above 3~MeV~\cite{csibkg} in the prototype measurements
suggests that a $^{238}$U and $^{232}$Th
concentration (assuming equilibrium)
of $< 10^{-12}$~g/g can be achieved.

\begin{figure}
\centerline{
\epsfig{file=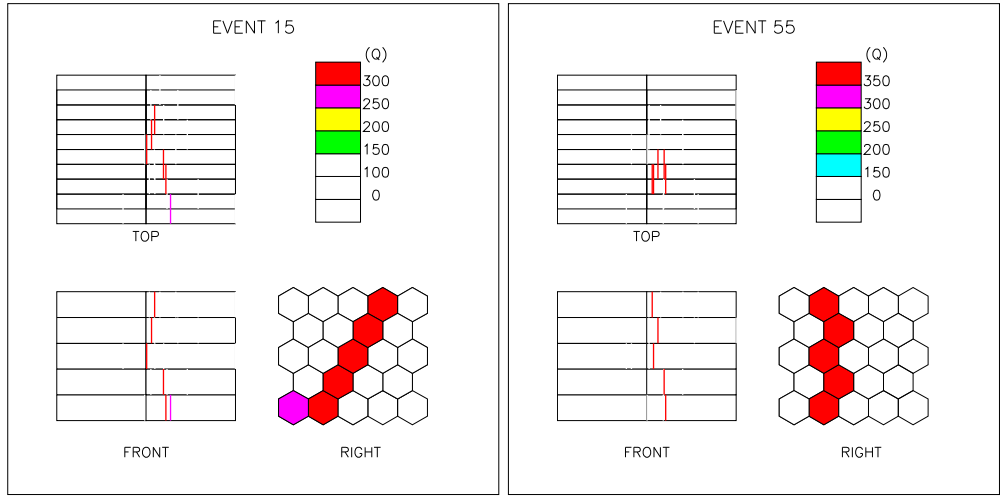,width=7cm,angle=90}
}
\caption{
Two typical cosmic ray events taken on
site at KS Lab with the
CsI(Tl) detector system.
}
\label{csimuon}
\end{figure}

The data taken from the CsI(Tl) detector for Period I 
would be used for further optimization of the operation
parameters as well as for studying the background.
A cosmic muon event is shown in Figure~\ref{csimuon}.
A $>$150~kg system will be installed for Period II. 
The physics goals include studies of neutrino-electron
and neutrino-nuclei scattering cross sections.

\section{R\&D Program}

Various projects with
stronger R\&D flavors are
proceeding in parallel to the 
reactor experiment. The highlights are :

\subsection{Low Energy Neutrino Detection}

It is recognized recently that $^{176}$Yb and $^{160}$Gd are
good candidate targets in the detection of
solar neutrino ($\nue$) by providing a flavor-specific
time-delayed tag\cite{lens}.
Our work
on the Gd-loaded scintillating crystal GSO\cite{gso}
indicated major background issues to be
addressed.
We are exploring the possibilities
of developing Yb-based scintillating crystals, like
doping the known 
crystals $\rm{Yb Al_{l5} O_{12}}$(YbAG) and
$\rm{Yb Al O_{3}}$(YbAP) 
with scintillators.

In addition, we have completed a feasibility study
on boron-loaded liquid scintillator for the detector
of $\nuebar$\cite{liqscin}. The case of 
``Ultra Low-Energy'' HPGe detectors,
with the potential applications of
Dark Matter searches  neutrino-nuclei coherent scatterings, 
are now being investigated. 

\subsection{Dark Matter Searches with CsI(Tl)}

\begin{figure}
\centerline{
\epsfig{file=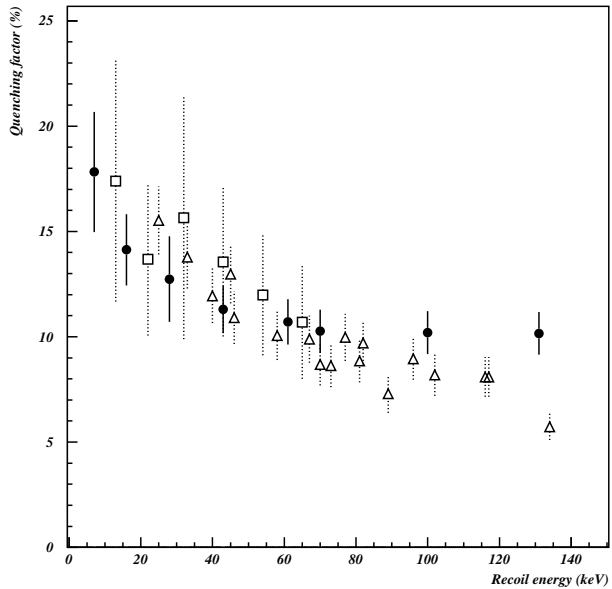,width=9cm}
}
\caption{
The quenching factors, shown as black circles,
measured at IAE Tandem, as compared to 
Open triangles and open squares from
previous work.
}
\label{csiqf}
\end{figure}

Experiments based on the mass range of 100~kg of NaI(Tl)
are producing some of the most sensitive results in
Dark Matter ``WIMP'' searches\cite{naicdm}.
The feasibilities
and technical details of adapting CsI(Tl) or other
good candidate crystal like CaF$_2$(Eu) for WIMP Searches
have been studied.
A neutron test beam measurement
was successfully performed at IAE 13~MV Tandem
accelerator\cite{nbeam}.
We have collected the world-lowest threshold data
for nuclear recoils in CsI, enabling us to
derive the quenching factors, displayed
in Figure~\ref{csiqf}, as well as
to study the pulse shape discrimination
techniques at the realistically low light output regime. 
The KIMS Collaboration 
will pursue such an experiment in South Korea\cite{kims}.

\subsection{Radio-purity Measurements with Accelerator Mass Spectrometry}

Measuring the radio-purity of detector target materials
as well as other laboratory components are crucial to
the success of low-background experiments.
The typical methods are direct photon counting with
high-purity germanium detectors, $\alpha$-counting
with silicon detectors or the neutron
activation techniques.
We are exploring the capabilities
of radio-purity measurements further with the
new Accelerator Mass Spectroscopy (AMS) techniques\cite{amsradio}.
This approach may be complementary to existing
methods since it is in principle a superior
and more versatile method as demonstrated in
the $^{13}$C system, and it is
sensitive to radioactive isotopes that do
not emit $\gamma$-rays (like single beta-decays
from $^{87}$Rb and $^{129}$I) or where $\gamma$
emissions are suppressed (for instance,
measuring $^{39}$K provides a gain of
10$^5$ in sensitivity relative to detecting
$\gamma$'s from $^{40}$K).
A pilot measurement of the $^{129}$I/$^{127}$I ratio
($< 10^{-12}$) 
in CsI was successfully performed
demonstrating the capabilities
of the Collaboration. Further beam time
is scheduled at the IAE AMS facilities\cite{ciaeams}
to devise measuring schemes for the other
other candidate isotopes like
$^{238}$U, $^{232}$Th, $^{87}$Rb, $^{40}$K 
in liquid and crystal scintillators
beyond the present capabilities by the
other techniques.

\subsection{Upgrade of FADC for LEPS Experiment}

\begin{figure}
\centerline{
\epsfig{file=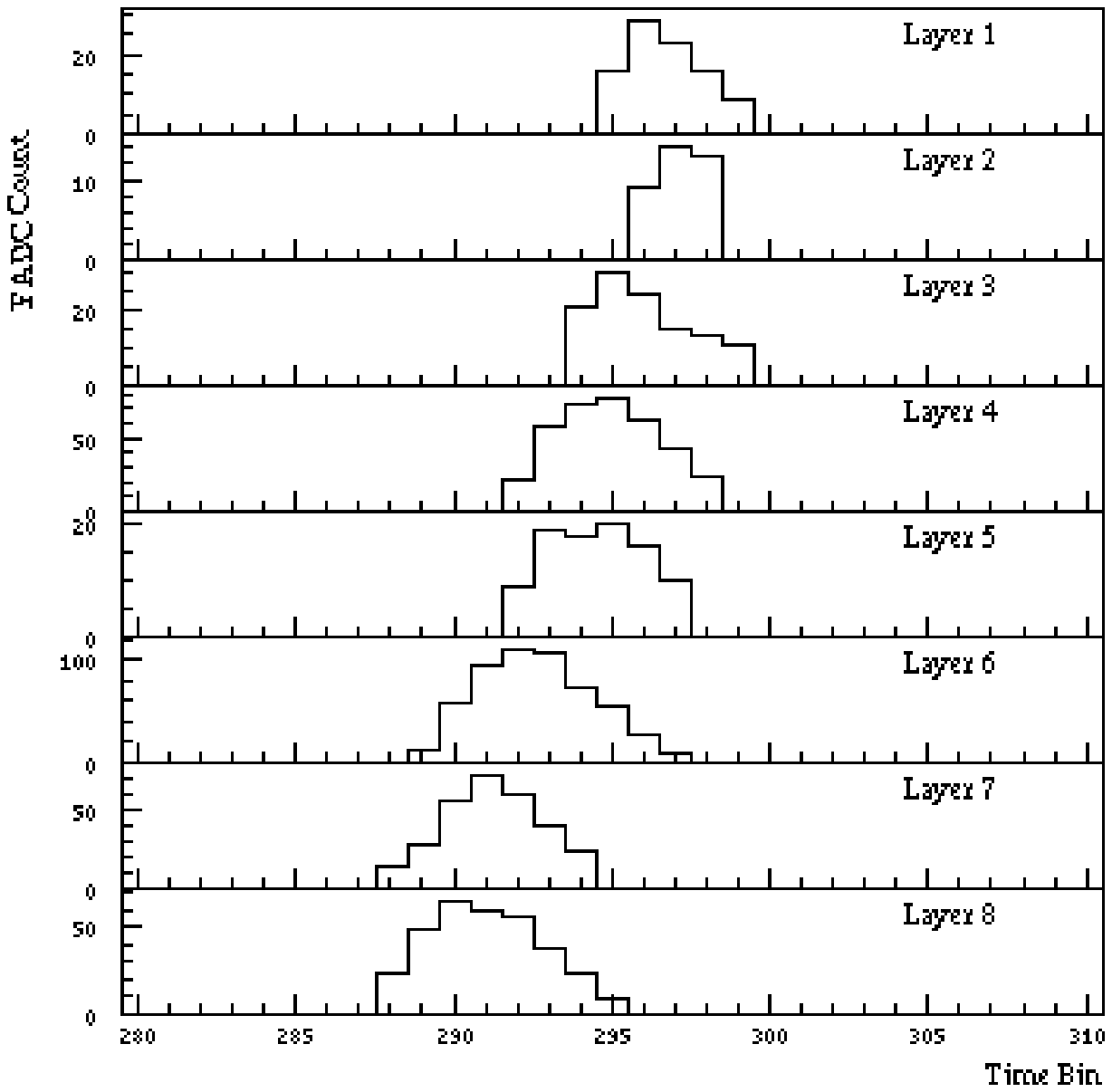,width=7cm}
\epsfig{file=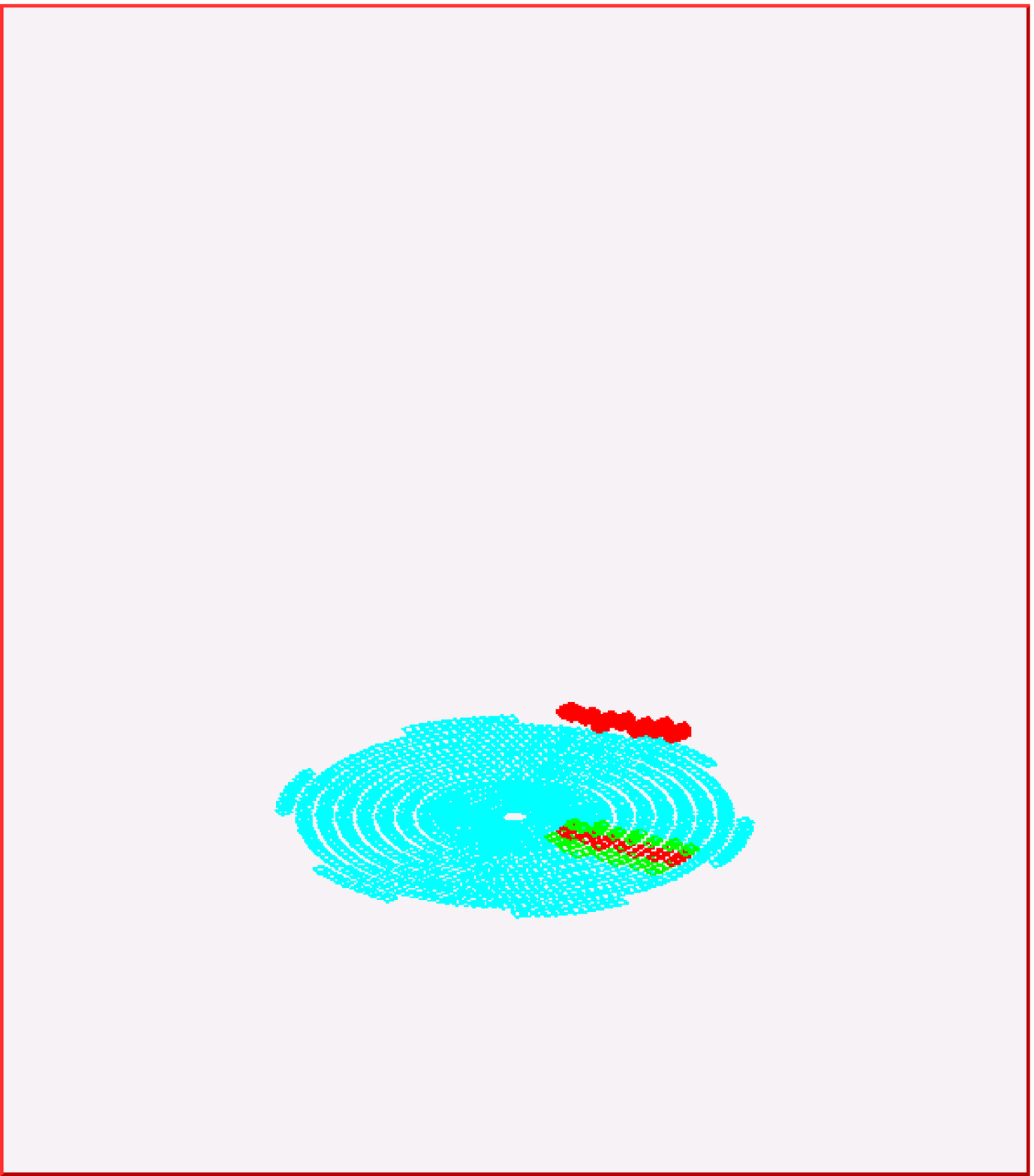,width=6cm}
}
\caption{
Measurements from the prototype TPC for LEPS Experiment,
with the TEXONO FADC system. Only one sector of of 
the TPC is equipped with readout electronics.
}
\label{fadctpc}
\end{figure}

Following the success in the design and operation
of the FADCs at the KS Lab, we will develop new FADCs
for the Time Projection Chamber (TPC)
constructed as a sub-detector for
the LEPS experiment at the SPring-8 Synchrotron
Facilities in Japan\cite{Spring8}.
The current FADCs are being used to provide readout
to test the prototype TPC, an event of
which is depicted in Figure~\ref{fadctpc}.
The upgraded FADCs
will have 40~MHz sampling rate, 10-bit dynamic
range  and be equipped with Field Programmable Gate Array (FPGA)
capabilities for real time data processing.
This new system is
expected to be commissioned in Fall 2002.

In addition, the Collaboration is participating in
the discussions on the scientific program and technical
feasibilities of (a) the ``H2B'' project\cite{h2b}: a
2000~km Very Long Baseline High Energy Neutrino
Experiment at Beijing to receive a neutrino
beam from the HIPA Facilities in Tokyo due
to be commissioned by 2006 in Japan, and
(b) the detection scheme of very high energy tau-neutrinos
using mountain ranges as target and air as the subsequent
showering volume\cite{uhenu}.

\section{Outlook}

With the strong evidence d
neutrino oscillations from atmospheric and
solar neutrino experiments\cite{conf}, there  are intense
world-wide efforts to pursue the next-generation
of neutrino projects. 
Neutrino physics and astrophysics will remain
a central subject in experimental particle physics
in the coming decade and beyond. There
are room for ground-breaking technical innovations -
as well as potentials for surprises in the scientific
results.

A Taiwan, China and U.S.A.
collaboration has been built up 
with the goal of establishing
a qualified experimental program
in neutrino and astro-particle physics.
It is 
the first generation collaborative efforts
in large-scale basic research between scientists
from Taiwan and Mainland China.
The technical strength and scientific
connections of the Collaboration
are expanding and consolidating.
The flagship experiment is to
perform the first-ever particle
physics experiment in Taiwan
at the Kuo-Sheng Reactor Plant.
From the Period I data taking, we expect to achieve
world-level sensitivities and neutrino magnetic
moments and radiative lifetime studies.
A wide spectrum of R\&D projects are 
being pursued.
New ideas are being explored within a bigger
framework.

The importance of the implications and 
outcomes of the experiment and 
experience will
lie besides, if not beyond, neutrino physics.

\section{Acknowledgments}

The authors are grateful to the scientific members,
technical staff and industrial partners
of TEXONO Collaboration, as well as 
the concerned colleagues
for the many contributions which ``make 
things happen'' in such a short period
of time.
Funding are
provided by the National Science Council, Taiwan
and 
the National Science Foundation, China, as well
as from the operational funds of the 
collaborating institutes.


\end{document}